\documentstyle[prl,aps,epsfig,twocolumn]{revtex}

\begin{document}

\title{Microscopic conditions favoring itinerant ferromagnetism: Hund's rule 
coupling and orbital degeneracy}
\author{K.~Held and D.~Vollhardt}
\address{Theoretische Physik III, Elektronische Korrelationen und
Magnetismus, Universit\"at Augsburg, D-86135 Augsburg, Germany}
\date{\today}

\maketitle

\abstract{
The importance of Hund's rule coupling for the stabilization of 
itinerant ferromagnetism
is investigated within a two-band Hubbard model.
The magnetic phase diagram is calculated
by finite-temperature quantum Monte Carlo simulations 
within the dynamical mean-field theory.
Ferromagnetism is found in a broad range of electron fillings 
whereas antiferromagnetism exists only near half filling.
The possibility of orbital ordering at quarter filling 
is also analyzed.}

\section{Introduction}
Investigations of the microscopic mechanisms responsible for the stability of
metallic ferromagnetism have recently received renewed attention. In
particular, it was finally established that itinerant ferromagnetism is
indeed stable in the one-band Hubbard model at intermediate 
on-site interactions, an important condition being a properly tuned kinetic
energy with a pronounced peak in the density of states (DOS) near the band 
edge \cite{MT,MuellerHartmann95,Hanisch97,Daul97,Hlubina97,Vollhardt97b,Ulmke98,Wahle98,Obermeier97}.
For real ferromagnets, e.g., the
transition metals Fe, Co, and Ni, 
quite a different microscopic mechanism  for ferromagnetism, based on
Hund's rule coupling in the presence of orbital degeneracy,
is also expected to be relevant. It was first suggested by
Slater \cite{Slater36} and then stressed  by van Vleck \cite{vanVleck53}
that the intra-atomic exchange leading to  
``Hund's rule atomic magnetism'' might be responsible 
for  bulk ferromagnetism, i.e., the hopping of
electrons or holes might lead to a bulk ordering of preformed atomic moments.

Concrete calculations based on microscopic models began with Roth \cite{Roth66}
who considered a two-band extension of the Hubbard model
\begin{eqnarray}
\hat{H} &=& -t \sum_{ i j \nu \sigma}  \hat{c}^{\dagger}_{i \nu 
\sigma} \hat{c}^{\phantom{\dagger}}_{j \nu \sigma} 
+ U \sum_{i \nu}  \hat{n}_{i\nu\uparrow}\hat{n}_{i\nu\downarrow} \nonumber \\
&&+  \! \sum_{i;\nu < \nu';\sigma\sigma'}  \! (V_0-\delta_{\sigma \sigma'}F_0)
  \hat{n}_{i\nu\sigma}\hat{n}_{i\nu'\sigma'} \nonumber\\&&
-F_0 \!  \sum_{i;\nu < \nu';\sigma \neq \sigma'}  \! 
 \hat{c}^\dagger_{i \nu \sigma} \hat{c}^{\phantom{\dagger}}_{i \nu \sigma'} 
 \hat{c}^\dagger_{i \nu' \sigma'} \hat{c}^{\phantom{\dagger}}_{i \nu' \sigma} 
\label{hm}
\end{eqnarray}
where $\nu$ denotes  the two (or more) orbitals, $\sigma$ 
the spin, and $i$, $j$ the lattice sites. 
In this model 
all interactions are purely local, i.e., occur only on a single site.
Apart from a Hubbard interaction $U$ for electrons of opposite spin
on the same orbital, there is also a density-density interaction $V_0$
between electrons on different orbitals, as well as an intra-atomic
exchange interaction $F_0$ which is separated into its
density-density and spin-flip contribution (the spin-flip term 
is neglected in Sec. \ref{rd}). An ``on-site pair hopping''-term of 
the same size is not considered since it requires an empty and a 
doubly-occupied orbital to take effect. Such configurations are 
strongly suppressed by the Hubbard interaction $U$ and the kinetic energy.
The kinetic energy describes the hopping of
electrons of a given spin between identical orbitals on different sites.

It is instructive to consider the two-site model
with one electron per atom, i.e., at quarter filling. 
In the strong-coupling regime the ground state is a
spin triplet and orbital singlet, i.e., on the two atoms 
{\em different} orbitals are occupied.
 Within second order perturbation theory the energy of this state is
readily calculated as $-4 t^2/(V_0-F_0)$. This  connection between
{\em staggered orbital ordering}
 and {\em ferromagnetism} was investigated already in 1966
 by Roth \cite{Roth66}:
applying 
the random phase (or Hartree-Fock) approximation to (\ref{hm})
she observed that,
 at $T=0$ and  quarter filling, 
a ferromagnetic state is unstable against an additional
staggered orbital ordering for $V_0-U/2-F_0/2>0$.
For decreasing temperatures the Hartree-Fock approximation
predicts  
first a phase transition from paramagnetic to 
ferromagnetic ordering, and at a lower temperature a second
transition to a phase with ferromagnetic and orbital ordering.
We note that the phenomenon of staggered orbital ordering may be a 
characteristic feature of any effective two-band model with orbital degeneracy
(as realized, for example, in $\mathrm e_g$-bands).

A major step towards the  understanding of the physics of the two-band model
was the derivation of an  effective strong-coupling Hamiltonian
at quarter filling by
Kugel' and Khomski\u{i} \cite{Kugel73}, and
by
Cyrot and Lyon-Caen \cite{Cyrot75}
 who included the effect of on-site pair hopping. This effective Hamiltonian
has coupled spin and orbital (pseudo spin) degrees of freedom.
Within a self-consistent-field approximation \cite{Kugel73},
where orbital and spin degrees of freedom are decoupled,
and also within the molecular field theory \cite{Cyrot75},
where correlations between different lattice sites are neglected,
the effective Hamiltonian shows an insulating ferromagnetic ground state with
staggered orbital ordering. Contrary to the 
weak-coupling  Hartree-Fock approximation this 
strong-coupling approach predicts that orbital ordering 
occurs first when the temperature is decreased \cite{Cyrot75}.
The ferromagnetic ground state at quarter filling
and sufficiently strong Coulomb interactions 
was confirmed by exact diagonalization studies
of finite systems in one dimension \cite{Gill87,Kuei97,Hirsch97}.

Off quarter or half filling  the high degeneracy in the atomic limit
makes a  perturbational analysis
essentially impossible \cite{Chao77}.  
Nevertheless, for infinite Coulomb interaction  $U=\infty$ and
one dimension,
M\"uller-Hartmann \cite{MuellerHartmann95} proved
 that for $F_0 > 0$ the ground state of (\ref{hm}) 
is ferromagnetic for fillings $0<n<2$.  
Recently exact diagonalization studies in one dimension  were
performed by Hirsch \cite{Hirsch97}.
The results  depend sensitively on the boundary conditions and the
number of lattice sites: below quarter filling  and for six 
lattice sites
ferromagnetism is found for the system with open boundary conditions, 
while it is absent when periodic boundary conditions are used.
Insight was also obtained by several approximative treatments, 
in particular the
 Hartree-Fock theory \cite{Cyrot75,Oles83},
a generalized Hartree-Fock  approach \cite{Fleck97}, 
more complex variational
wave functions \cite{Okabe97}, the local approach \cite{Olesetc},
and the Gutzwiller 
approximation \cite{Buenemann}.
They all find ferromagnetism to be stabilized by Hund's rule coupling
at  intermediate to strong Coulomb interactions.
Clearly, in this regime a proper treatment of correlation effects 
and the dynamics of the quantum mechanical many-body problem
is essential.
In the last few years the dynamical mean-field theory  (DMFT)
has turned out to provide a reliable framework and powerful method
for the investigation of such types of problems.
Hence we use it in this paper to study the stability of the 
ferromagnetic phase at and off quarter filling for intermediate
values of the Coulomb interaction, and to determine transition 
temperatures.

The paper is structured as follows.
In Sec.~\ref{qmc} the quantum Monte Carlo (QMC) algorithm  to
solve the DMFT equations is introduced. The 
magnetic phase diagram is presented in Sec.~\ref{fm}, and the possibility 
of orbital ordering is discussed in Sec.~\ref{oo}.

\section{Quantum Monte Carlo solution of the dynamical mean-field equations}
\label{qmc}
The DMFT \cite{Metzner89a,MuellerHartmann89a,Janis,Jarrell92,Georges92a,note3,Georges96} approximates the lattice model by a single-site
problem of electrons in an effective medium (mean-field) that may
be described by a frequency dependent, i.e., dynamical, self-energy 
$\Sigma_{\nu \sigma} (\omega)$. The latter has to be determined 
self-consistently
via a $\mathbf k$-integrated Dyson equation that reads 
\begin{equation}
G_{\nu \sigma n} = 
\int\limits_{-\infty}^\infty d \varepsilon \,\frac{N^0 (\varepsilon)}
{i \omega_n +  \mu - \Sigma_{\nu \sigma n}- \varepsilon}
\label{dyson}
\end{equation}
with Matsubara frequencies $\omega_n$, Green function 
$G_{\nu \sigma  n}=G_{\nu \sigma }(i \omega_n)$, 
and the DOS of the non-interacting electrons $N^0(\epsilon)$.
In the case of the multi-band Hubbard model (\ref{hm}) 
the single-site problem takes the form
\begin{equation}
G_{\nu \sigma n} = - \frac{1}{\cal Z}
 \int {\cal D} [\psi ] {\cal D} [\psi^*]  
\psi_{\nu \sigma n}^{\phantom *} \psi_{\nu \sigma n}^*
 e^{{\cal A} [ \psi, \psi^*,{\cal G}^{-1}]}
\label{siam}
\end{equation}
where $\psi^{\phantom *}$ and $\psi^*$ are Grassmann variables,
${\cal G}^{-1}=G^{-1}+\Sigma$, and 
${\cal A} [ \psi, \psi^*,{\cal G}^{-1}]$ denotes 
the single-site action including all local interactions
\begin{eqnarray}
\lefteqn{{\cal A} [\psi,\psi^*,{\cal G}^{-1}]= \sum_{\nu \sigma, n} 
\psi_{\nu \sigma n}^*
{\cal G}^{-1}_{\nu \sigma n} \psi_{\nu \sigma n}^{\phantom *}}
\label{action}\\&&
- U  \sum_{\nu} \int\limits_0^\beta
d \tau \, \psi_{\nu\uparrow}^* (\tau) \psi_{\nu \uparrow}^{\phantom *} (\tau)
 \psi_{\nu \downarrow}^* (\tau) \psi_{\nu \downarrow}^{\phantom *} (\tau)
\nonumber\\&&
-  \! \sum_{\nu<\nu';\sigma \sigma'} \! (V_0-\delta_{\sigma \sigma'}F_0)  \!
\int\limits_0^\beta  \!
d \tau \, \psi_{\nu\sigma}^* (\tau) \psi_{\nu \sigma}^{\phantom *} (\tau)
 \psi_{\nu' \sigma'}^* (\tau) \psi_{\nu' \sigma'}^{\phantom *} (\tau)
\nonumber\\&&
+ F_0 \! \sum_{\nu<\nu';\sigma \neq \sigma'}
\int\limits_0^\beta 
d \tau  \, \psi_{\nu\sigma}^* (\tau) \psi_{\nu \sigma'}^{\phantom *} (\tau)
 \psi_{\nu' \sigma'}^* (\tau) \psi_{\nu' \sigma}^{\phantom *} (\tau). \nonumber
\end{eqnarray}
As for the one-band model \cite{Jarrell92,Georges92a} Eqn.~(\ref{siam})
is equivalent to a (degenerate) Anderson impurity model and can thus
be treated by standard techniques. In the present paper the 
QMC algorithm of Hirsch and Fye \cite{Hirsch86} will be used. 
In a first step each one of the four terms of Eqn.~(\ref{action}) are 
decomposed via the Trotter-Suzuki formula, and  imaginary time
is discretized ($\tau = l \, \Delta \tau$ for
$l=1 \ldots \Lambda$ and $\Delta \tau=\beta/\Lambda$).
In the second step the interaction terms are decoupled
to obtain a quadratic action.
For the density-density interactions this is achieved as usual by the
Hubbard-Stratonovich transformation
\begin{eqnarray}
 \exp \{\frac{\Delta\tau}{2} (V_0-F_0)(\psi_{\nu \sigma l}^*
 \psi_{\nu \sigma l}^{\phantom *}-\psi_{\nu' \sigma l}^*
 \psi_{\nu' \sigma l}^{\phantom *})^2 \}
= &&\\  \nonumber \frac{1}{2}
 \sum_{s  = \pm 1}\exp \{ \lambda s (\psi_{\nu \sigma l}^*
 \psi_{\nu \sigma l}^{\phantom *}-\psi_{\nu' \sigma l}^*
 \psi_{\nu' \sigma l}^{\phantom *})\},&&
\end{eqnarray}
where $\cosh(\lambda)=\exp(\Delta\tau (V_0-F_0)/2)$. 
The same decoupling holds for the remaining $V_0$ and the $U$ terms.
For $N$ orbitals this yields altogether
$(2N^2-N)\Lambda$ auxiliary fields for the density-density interactions.

By contrast, there does not exist  a standard decoupling
scheme for 
the last term of Eqn.~(\ref{action})
describing a spin-flip. 
Recently  Motome and Imada \cite{Motome97} proposed 
a decoupling scheme for this term: it avoids the minus-sign problem in the 
symmetric case at half filling, but leads to a phase problem off half filling
due to the use of a complex auxiliary field.
Instead, we tried the following Hubbard-Stratonovich decoupling:
\begin{eqnarray}
 \lefteqn{\exp ( \Delta\tau F_0 \psi_{\nu \sigma l}^*
 \psi_{\nu -\sigma l}^{\phantom *}\psi_{\nu' -\sigma l}^*
 \psi_{\nu' \sigma l}^{\phantom *} )
=}&&\\&&\nonumber \frac{1}{2}
 \sum_{s  = \pm 1}\exp \{     \mu s (\psi_{\nu \sigma l}^*
 \psi_{\nu -\sigma l}^{\phantom *}+\psi_{\nu' -\sigma l}^*
 \psi_{\nu' \sigma l}^{\phantom *})\},
\end{eqnarray}
with $\mu = \sqrt{\Delta \tau \, F_0}$. Unfortunately, this transformation
was found to 
lead to a sign-problem, too, even within the DMFT where the sign-problem
is absent in the single-band case. For this reason we neglect the 
spin-flip term in the following.
Hund's rule coupling is thus restricted to the direction of the
quantization axis, i.e., the $z$-axis, implying the breaking of the SU(2)
spin rotation symmetry.
The restriction to an Ising-type Hund's rule coupling
is commonly used in the investigations 
of the two-band Hubbard model \cite{Roth66,Cyrot75,Oles83,Fleck97}.
This restriction has no effect on the critical temperatures 
in the limits of weak and strong coupling, i.e., 
in the Hartree-Fock approximation and the Weiss mean-field theory
of the corresponding spin model, respectively. Therefore we expect 
that the spin-flip term has no strong effect on the critical temperature
at intermediate coupling. Note, that while the influence of the
spin-flip term on  critical temperatures is probably small
the excitation spectrum at T=0 depends sensitively on this term.
It shows a spin gap or not, respectively.

Based on the above Hubbard-Stratonovich decoupling 
 the Monte Carlo method is
employed to sample the auxiliary spin configurations and thus to
calculate Green functions and susceptibilities. The multi-band
algorithm is a generalization of the  one-band algorithm.
A similar generalization was employed by Rozenberg \cite{Rozenberg97} 
in the investigation
of metal-insulator transitions in the two-band Hubbard model without Hund's
rule coupling.

Phase boundaries are determined by a Curie-Weiss fit of the 
corresponding susceptibility in the homogeneous phase. 
Since the corresponding critical temperatures 
still depend on the unphysical time discretization parameter $\Delta \tau$
a second order fit to $\Delta \tau=0$ was performed from at least six
values of $\Delta \tau \in [0.075,0.5]$. Besides the statistical error
of the QMC simulation (propagated via $T$- and $\Delta \tau$-fit), there 
exists an additional systematic error due to higher order contributions in the
 $T$ and $\Delta \tau$ fits. In particular the $\Delta \tau$ 
dependence of the Curie temperature was considerable for the
data of Sec.~\ref{rd}. Therefore we estimated
this systematic 
error by  comparing the Curie temperature obtained
from all $\Delta \tau$ values to that calculated 
without the $\Delta \tau=0.5$ value.
The  individual and mean 
difference  between these two fits was within
the statistical error of the Monte Carlo data.
This analysis implies that the systematic error is  smaller
than the statistical error shown in the figures.

\section{Results and Discussion}
\label{rd}
\subsection{Ferromagnetism}
\label{fm}
In the present paper we investigate the 
Hund's rule coupling in the presence of orbital degeneracy
as a possible origin of itinerant ferromagnetism.
This microscopic mechanism should be distinguished from 
the one found to be important for the
 single-band Hubbard model, which is based on an 
asymmetric DOS   \cite{MT,MuellerHartmann95,Hanisch97,Daul97,Hlubina97,Vollhardt97b,Ulmke98,Wahle98,Obermeier97}.
Therefore
we employ a  symmetric semielliptic DOS
$N^0(\epsilon) = \sqrt{(2t^*)^2 \! -\! \epsilon^2 }/(2 \pi t^{*2})$
(in the following $t^*\equiv 1$ will set our energy scale).
For this DOS and a symmetric Gaussian DOS
no ferromagnetism was found in the 
single-band Hubbard model up to a
 Coulomb interaction of $U=30$  \cite{Wahle98,Obermeier97}. 

\begin{figure}
\unitlength1cm \epsfxsize=8.6cm \hspace{3.5cm}  \epsfbox{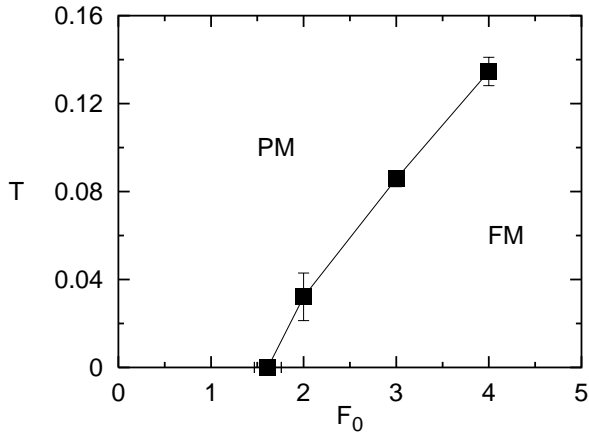}
\caption{Curie temperature vs.~Hund's rule coupling $F_0$ for the two-band 
Hubbard model at $U=9$, $V_0=5$, and $n=1.25$. 
Ferromagnetism is seen to be stabilized by Hund's rule coupling.}
\label{TvsF}
\end{figure}

In fact, even the  two-band model {\em without} Hund's
rule coupling does not indicate ferromagnetism at $U=9$, $V_0=5$, and 
$n=1.25$ (see Fig.~\ref{TvsF}).
However, already a small Hund's rule coupling, $F_0= 1.61 \pm 0.15$, 
is sufficient to stabilize  ferromagnetism \cite{note2}.
This value is considerably smaller than  that obtained by Kuei
and Scalettar \cite{Kuei97}: in one dimension (six sites)
at  $n=1$ with $U=V_0+F_0$ no ferromagnetism was found 
below $F_0\approx 8$ (in units rescaled to obtain
a bandwidth of four). This value of $F_0$ is, however, comparable 
to the
\begin{figure}[t]
\unitlength1cm \epsfxsize=8.6cm \hspace{3.5cm}  \epsfbox{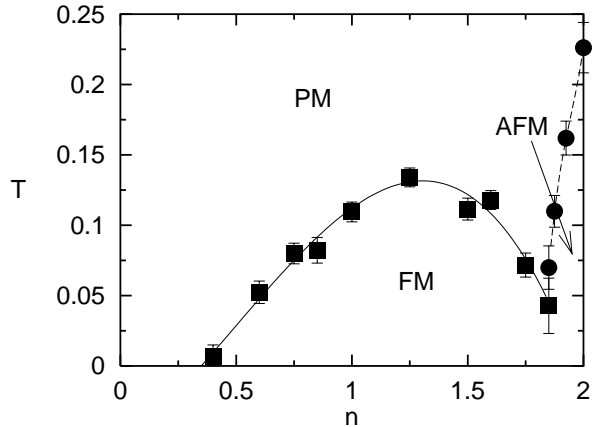}
\caption{Magnetic $T-n$ phase diagram for $U=9$, $V_0=5$, and $F_0=4$
with paramagnetic (PM), ferromagnetic (FM), and antiferromagnetic phase (AFM).
 Ferromagnetism is found in a broad range of fillings.}
\label{TvsN}
\end{figure}
\begin{figure}
\vspace{-.5cm}
\unitlength1cm \epsfxsize=8.6cm \hspace{3.5cm}  \epsfbox{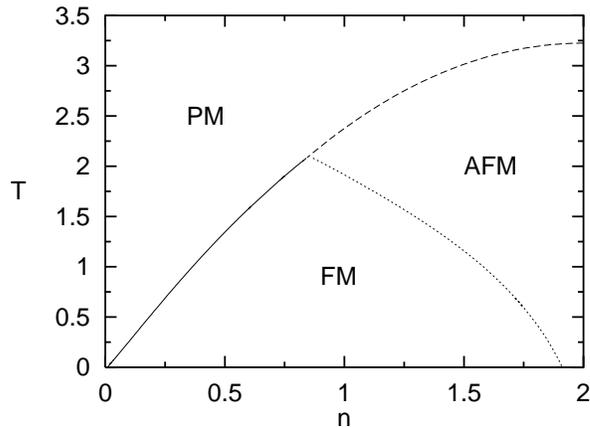}
\caption{Same as Fig.~\ref{TvsN} but calculated 
within the Hartree-Fock approximation
(solid line: Curie temperature, dashed line: N{\'e}el temperature, dotted line:
 first order transition between FM and AFM).  
}
\label{TvsNHF}
\end{figure}
results of Hirsch \cite{Hirsch97}:  in one dimension
at  $n=1$ and $n=1.5$ (six and four sites, respectively) 
with $U=V_0+2 F_0$ no 
ferromagnetism was found below $F_0\approx 1.5$ and  $2$, respectively.
Note, that the relation  $U=V_0+F_0$ makes the
Hamiltonian (\ref{hm}) form-invariant w.r.t. orbital rotations, 
while  this holds for $U=V_0 + 2 F_0$ if an additional pair hopping
term is added. 
The Hund's rule coupling $F_0$ must be smaller than the
density-density interaction $V_0$ since 
otherwise an unphysical
attractive interaction between two electrons
on the same site
(in different orbitals with the same spin) exists.

The magnetic $T - n$ phase diagram is presented in Fig.~\ref{TvsN}
for $U=9$, $V_0=5$, and a relatively strong Hund's rule coupling
$F_0=4$. 
Ferromagnetism is found in a broad range of electron
densities $n$. Note, that the phase-diagram is symmetric
around $n=2$ due to particle-hole symmetry. 
The special case of quarter filling $n=1$ does not mark
a pronounced point in the phase diagram. In particular the maximal Curie 
temperature is found {\em above} quarter filling. 
Typical Curie temperatures are 
about $0.1$, which for a bandwidth of $4$eV ($t^* \equiv$ 1eV)  corresponds 
to $0.1$eV, i.e., about
$1000$K. Near half filling the 
antiferromagnetic Heisenberg exchange suppresses the ferromagnetic order 
and a narrow antiferromagnetic phase with the usual AB sublattice structure develops. 
Fig.~\ref{TvsN} suggests that Curie and N\'eel temperature cross
at a finite temperature and not at $T=0$. This is confirmed
by the observation that for $n=1.85$ the Curie temperature
extrapolated from the paramagnetic phase lies only slightly below the
N\'eel temperature.

The same $T - n$ phase diagram calculated
within the Hartree-Fock approximation  is shown in Fig.~\ref{TvsNHF}. 
The Hartree-Fock 
approximation  fails to describe the suppression of the 
magnetic order at the crossover from ferro- to antiferromagnetism.
Furthermore, the magnetic phases
are overestimated, i.e., the critical temperatures are more than an order of 
magnitude too large and  both magnetic phases
 continue to extremely small values of $n$. Within the Hartree-Fock 
approximation a first order phase transition between antiferromagnetic 
and ferromagnetic phase occurs. It is not clear at present whether this 
is also true within the DMFT.

\subsection{Orbital ordering}
\label{oo}
\begin{figure}
\unitlength1cm \epsfxsize=8.6cm \hspace{3.5cm}  \epsfbox{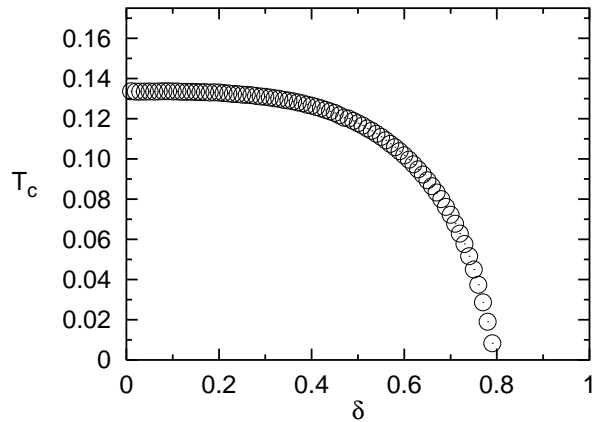}
\caption{Critical temperature for the instability 
against a state with mixed ferromagnetic and orbital ordering 
(control parameter $\delta$)
as obtained from $[\chi(T_c,\delta)]^{-1}=0$. For $U=9$, $V_0=5$, $F_0=4$, 
and $n=1$
the highest temperature is 
found at $\delta=0$, indicating a transition into a purely ferromagnetic state.
}\label{oo1}
\end{figure}

As was pointed out in the introduction, at quarter filling 
the Weiss mean-field 
theory for the effective Kugel' and Khomski\u{i} Hamiltonian 
\cite{Kugel73,Cyrot75}
predicts an instability of the paramagnetic phase 
against staggered orbital ordering when the temperature is decreased,
while a transition to  
pure ferromagnetic order is suggested by the Hartree-Fock approximation.
The DMFT, containing both approximations as limits at
strong and weak coupling, 
respectively, is well suited to clarify this contradiction. 
To investigate a possible transition from the 
phase without long-range order to a phase with {\em mixed} ferromagnetic and
staggered orbital ordering we introduce a parameter $\delta$ that
allows one to investigate the instability w.r.t.~a general mixed ordering.
For every $\delta$ the corresponding susceptibility is defined through 
a field $h$, which 
modifies the grand potential 
$\Omega$ by the term
\begin{eqnarray}
\hat{H}_h &=& - h \sum_{i  \nu \sigma} \hat{n}_{i  \nu \sigma} 
\left\{ (1\!-\!\delta)\; \sigma  + \delta \;(-1)^i \nu \right\} \\
\chi  &=& - \frac{1}{L} \frac{\partial^2 \Omega}{\partial h^2}.
\end{eqnarray}
Here $L$ denotes the number of lattice sites.
For $\delta=0$ one obtains
the (para-) magnetic susceptibility and for $\delta=1$
the susceptibility for orbital ordering.
Fig.~\ref{oo1} shows that for  $U=9$, $V_0=5$, and $F_0=4$  the 
critical temperature is maximal for  $\delta=0$, i.e., the paramagnet
is unstable against pure ferromagnetic order. Thus, the phase diagram
Fig.~\ref{TvsN} need not be 
 modified  by phase transitions
from  paramagnetic  to  orbital ordering.
Whether the ferromagnetic phase becomes unstable 
against orbital ordering at even lower temperatures cannot be answered by the
method employed here, i.e., by the calculation of susceptibilities
within the 
paramagnetic phase. 

The scenario of Fig.~\ref{oo1} 
qualitatively agrees with the Hartree-Fock
approximation, which even at $T=0$ does not predict orbital ordering since
$V_0-U/2-F_0/2<0$. 
The disagreement with  Weiss mean-field 
theory can be explained by the fact

\begin{figure}
\vspace{-.5cm}
\unitlength1cm \epsfxsize=8.6cm \hspace{3.5cm}  \epsfbox{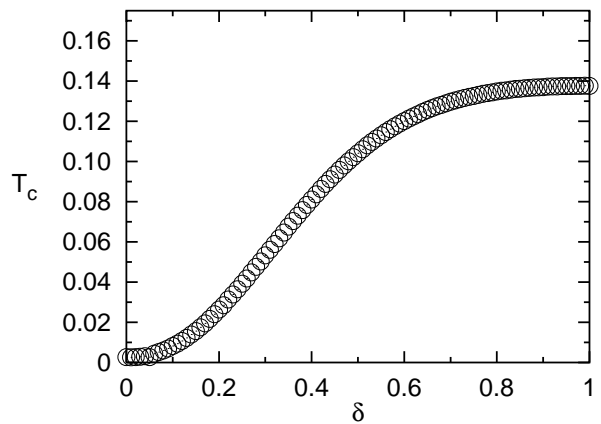}
\caption{Same as in Fig.~\ref{oo1} but for $U=8$, $V_0=6$, $F_0=2$, and $n=1$. 
Here, the maximal temperature is found at $\delta=1$, i.e.,
an instability against staggered orbital ordering occurs.
}\label{oo2}
\end{figure}
that the second order perturbation
theory leading to the effective strong-coupling Hamiltonian is
controlled in $t^2/(V_0-F_0)$ which is of ${\cal O} (1)$ here.

\begin{figure}[t]
\unitlength1cm \epsfxsize=8.6cm \hspace{3.5cm}  \epsfbox{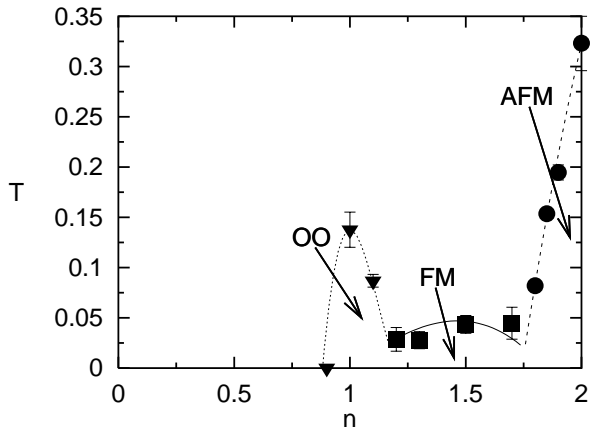}
\caption{$T-n$ phase diagram for $U=8$, $V_0=6$, and $F_0=2$. 
In addition to 
the phases in  Fig.~\ref{TvsN}  an  orbital ordering (OO) phase
is found near quarter filling.\label{pd2}}
\end{figure}

The considerations above suggest that orbital ordering may occur if the
intra-atomic exchange $F_0$ becomes smaller. Indeed,
for $U=8$, $V_0=6$, and $F_0=2$ a phase transition to
pure orbital ordering ($\delta =1$) occurs (see Fig.~\ref{oo2}). 
Fig.~\ref{pd2} shows the $T-n$ phase diagram for these parameters, 
including orbital ordering near quarter filling, antiferromagnetism
near half filling and ferromagnetism in between. 
We should mention, 
that although at $n=0.9$ the inverse orbital ordering susceptibility 
decreases with decreasing temperatures in a Curie-Weiss like behavior, suggesting
a transition at $T\approx 0.04$, it decreases again at lower temperatures. Hence
orbital ordering does not take place at  $n=0.9$.

Even at $n=1.2$, i.e., the
crossing point of the orbital ordering and ferromagnetic phase boundary
in Fig.~\ref{pd2},
no instability against a mixed ferromagnetic and orbital ordering is found.
Therefore we conclude that the two-band Hubbard model with Hund's rule coupling
$F_0$ shows an instability towards either pure ferromagnetic or pure
orbital ordering. However, since these
phases do not exclude each other a phase with mixed order may appear
at even lower temperatures, as is predicted by  Hartree-Fock and Weiss 
mean-field theory.

\subsection*{Conclusion}

We showed that even for a symmetric DOS
the Hund's rule coupling provides an
effective microscopic 
mechanism for the stabilization of ferromagnetism in a broad
range of electron densities off half filling.
This mechanism takes effect at intermediate to strong values of
the Coulomb interaction and therefore requires a proper treatment 
of the quantum mechanical correlations.
It is different
from the mechanism based on
an asymmetric DOS, which leads to ferromagnetism
even in the single-band Hubbard model. 
The question which one of these mechanisms
is the main driving force for ferromagnetism in Fe, Co, and Ni
remains open, if it can be explained by a single mechanism at all.
           
\subsubsection*{Acknowledgments}
The authors acknowledge helpful discussions with M. Ulmke, J.~Schlipf, 
and M.~Kollar. A generous computing account on the VPP 700 was 
provided by the LRZ M\"unchen.


\begin{thebibliography}{10}
\bibitem{note2}
The critical value of $F_0$ was determined by a linear interpolation between
  the Curie temperature for $F_0=2$ and the (negative)  Curie temperature
for $F_0=1$.
\end{thebibliography}
\end{document}